\documentclass[aps,amsmath,preprint,superscriptaddress,nofootinbib]{revtex4-1}  
\usepackage{graphicx}
\usepackage{changes}
\usepackage{color}
\usepackage{amsmath,amssymb}	
\usepackage{hyperref}
\usepackage{setspace}
\usepackage{comment}
\usepackage{array}  
\usepackage{cancel}
\usepackage{enumitem}
\usepackage{mathrsfs}
\usepackage{physics}
\usepackage{todonotes}
\usepackage{subcaption}
\usepackage{soul}

\begin{document}
\newcolumntype{Y}{>{\centering\arraybackslash}p{23pt}} 



\title{QCD Axion From a Spontaneously Broken B-L Gauge Symmetry}

\author{Gongjun Choi}
\email[e-mail: ]{gongjun.choi@gmail.com}
\affiliation{Tsung-Dao Lee Institute, Shanghai Jiao Tong University, Shanghai 200240, China}
\author{Motoo Suzuki}
\email[e-mail: ]{m0t@icrr.u-tokyo.ac.jp}
\affiliation{Tsung-Dao Lee Institute, Shanghai Jiao Tong University, Shanghai 200240, China}
\author{Tsutomu T. Yanagida}
\email[e-mail: ]{tsutomu.tyanagida@ipmu.jp}
\affiliation{Tsung-Dao Lee Institute, Shanghai Jiao Tong University, Shanghai 200240, China}
\affiliation{Kavli IPMU (WPI), UTIAS, The University of Tokyo, Kashiwa, Chiba 277-8583, Japan}
\date{\today}
\begin{abstract}

In this paper, we show that the Peccei-Quinn (PQ) symmetry with a good quality can be realized in a simple $B-L$ extension of the minimal supersymmetric standard model.
The PQ symmetry is a remnant of the $B-L$ gauge symmetry at the renormalizable level.
Besides, the sufficient quality of the PQ symmetry is preserved by a non anomalous discrete gauged $R$-symmetry and a small gravitino mass $m_{3/2}\ll 100$\,GeV.
 A viable mass range is 
 $m_{3/2}=\mathcal{O}(1)$\,eV
 which allows a high reheating temperature and many baryogenesis scenarios typified by the thermal leptogenesis without any astrophysical and cosmological problems. Such a light gravitino may be tested in the future 21cm line observations.

\end{abstract}

\maketitle


\newpage
\section{Introduction}
\label{sec:introduction}
A prime candidate for a solution of the Strong CP problem is to postulate a global chiral $U(1)$ symmetry called the Peccei-Quinn (PQ) symmetry, $U(1)_{PQ}$~\cite{Peccei:1977hh,Peccei:1977ur}. The PQ symmetry is exact at the classical level but anomalous with respect to the color gauge symmetry $SU(3)_c$. The main prediction of the PQ mechanism is the existence of the so-called axion which is a pseudo Nambu-Goldstone (NG) boson from the spontaneous breaking of $U(1)_{PQ}$~\cite{Weinberg:1977ma,Wilczek:1977pj}. Ongoing and future experiments shall look for the signal of the axion~(see $e.g.$ Refs.~\cite{Braine:2019fqb,Lakic:2020cin,Beurthey:2020yuq,Salemi:2019xgl,Alesini:2019nzq}).

Despite the success of the PQ mechanism, there remain two long-standing puzzles.
One problem is the so-called the axion quality problem. It is believed that all global symmetries should be broken by quantum gravity effects~\cite{Hawking:1987mz,Lavrelashvili:1987jg,Giddings:1988cx,Coleman:1988tj,Gilbert:1989nq,Banks:2010zn}.
Then, the explicit breaking of the PQ symmetry must be suppressed to an extraordinary degree, otherwise the PQ mechanism cannot explain the small QCD vacuum angle. The other issue is the origin of the PQ symmetry. The PQ symmetry may be realized as an accidental symmetry owing to some gauge symmetries like the baryon and lepton symmetries in the Standard Model (SM).

In this paper, we show that an accidental PQ symmetry can be realized by a simple extension of a model based on the $B-L$ gauge symmetry. The $B-L$ gauge symmetry leads to the emergence of the PQ symmetry at the classical level. To solve the quality problem, the model is also extended by the supersymmetry (SUSY), where the gauged discrete $R$-symmetry plays a role for enough suppression of explicit PQ breaking operators. One unique prediction of the model is the gravitino with small mass, $i.e.$ $m_{3/2}\ll 100$\,GeV. In particular, a gravitino mass range of  $m_{3/2}<\mathcal{O}(1)$\,eV is consistent with a high reheating temperature, allowing many baryogenesis scenarios typified by the thermal leptogenesis~\cite{Fukugita:1986hr} without suffering from neither astrophysical nor cosmological problems~\cite{Pagels:1981ke,Moroi:1993mb,Feng:2010ij}. This mass range of the gravitino mass may be searched by future observations of $21$cm line fluctuation~\cite{Oyama:2016lor}.

The organization of the paper is as follows. In Sec.~\ref{sec:model}, we propose an $B-L$ extension of the minimal supersymmetric standard model. In Sec.~\ref{sec:PQ}, we show the existence of the accidental PQ symmetry. In Sec.~\ref{sec:pheno}, we discuss the axion quality problem and several constraints on the model. The final section is devoted to our conclusions.

\section{The B-L gauge symmetry in a SUSY standard model}
\label{sec:model}
In this section, we discuss a $B-L$ extension of the minimal supersymmetric standard model (MSSM). 
In the following, we use the $SU(5)$ notation for presentational simplicity. But, we do not consider a full $SU(5)$ theory and do not introduce colored Higgs multiplets as seen below.

The $B-L$ gauge symmetry $U(1)_{B-L}$ is the most plausible extension to the SM. In the SM, the $B-L$ symmetry is realized as an accidental global symmetry. Gauging this symmetry requires additional $B-L$ charged fields from the gauge anomaly cancellation. Promising candidates are the three families of right-handed neutrinos. With this set-up, the smallness of the neutrino masses can be explained by the see-saw mechanism~\cite{Yanagida:1979as,GellMann:1980vs,Minkowski:1977sc} and the baryon asymmetry can be generated via the leptogenesis~\cite{Fukugita:1986hr} by the out-of-equilibrium decay of the right-handed neutrinos. 

Motivated by the above facts, we consider a model based on the $B-L$ extension of the supersymmetric standard model.%
\footnote{In Sec.~\ref{sec:pheno}, we will see that the supersymmetry is also motivated to protect the PQ symmetry from quantum gravity effect.}
For notational convenience, we use the so-called fiveness, $5(B-L)-4Y$ for the MSSM fields instead of $B-L$. The fiveness symmetry is realized as a linear combination of the $B-L$ gauge symmetry and $U(1)_Y$. The fiveness symmetry is intrinsically equivalent to the $B-L$ symmetry, and thus we call the fiveness $B-L$ from here on.
The $B-L$ charges of the chiral superfields in the MSSM and three right-handed neutrinos are
\begin{align}
{\bf 10}_{\rm SM}(+1),~\bar{\bf 5}_{\rm SM}(-3),~\bar N_R(+5)\ ,
\end{align}
where the MSSM matter fields are denoted by ${\bf 10}_{\rm SM}$ and $\bar{\bf 5}_{\rm SM}$ in the $SU(5)$ notation, $\bar N_R$ are the right handed neutrinos, and $(q)$ denote the $B-L$ charges.
Henceforth, we omit the flavor indices for simplicity. The two Higgs doublet supermultiplets $H_u$ and $H_d$ have $-2$ and $+2$ charges of $B-L$ (fiveness), respectively.

The majorana masses of the right handed neutrinos are provided when one introduces the SM gauge singlet chiral superfields,
\begin{align}
\Phi(-10),~\bar\Phi(10)\ ,
\end{align}
which couple to the right-handed neutrinos in the superpotential,
\begin{align}
W=y_N \Phi(-10)\bar N_R(+5)\bar N_R(+5)\ .
\end{align}
Here, $y_N$ denotes dimensionless coupling.%
\footnote{$\bar\Phi(10)$ is introduced for the anomaly free $B-L$ symmetry.}
The vacuum expectation values (VEVs) of $\Phi$ and $\bar\Phi$ are obtained by the superpotential
\begin{align}
W=X(2\Phi \bar\Phi-v^2)\ ,
\end{align}
where $X$ is a SM and $B-L$ gauge singlet chiral superfield, $v$ denotes a parameter with a mass dimension, and we are assuming the existence of the $R$-symmetry. $X$ has $R$-charge $+2$. $\Phi$ and $\bar\Phi$ have zero $R$-charges. We will discuss more details about the $R$-symmetry soon.

Based on the above gauged $B-L$ SUSY model, we may extend the model to solve the Strong CP problem. So far there is no PQ symmetry. However, it would be interesting to consider that a global PQ symmetry is a remnant of the $B-L$ gauge symmetry. To establish a model, we introduce pairs of the chiral superfields of ${\bf 5}$ and $\bar{\bf 5}$ representations of $SU(5)$~\cite{Kim:1979if,Shifman:1979if} and we assume that the pairs have nontrivial $B-L$ charges.
 Indeed, if the ${\bf 5}$ and $\bar{\bf 5}$ fields do not have $B-L$ charges, the operator ${\bf 5}\bar{\bf 5}$ is invariant under the $B-L$ gauge symmetry. Then, we do not obtain an anomalous PQ symmetry because only a vector-like symmetry is allowed.%
\footnote{See Ref.~\cite{Harigaya:2013vja} for the accidental PQ symmetry by a gauged discrete $R$-symmetry.}
Besides, even if the ${\bf 5}$ and $\bar{\bf 5}$ fields have opposite $B-L$ charge, the operator ${\bf 5}\bar{\bf 5}$ is still allowed. Therefore, we consider the charge assignments where the mass terms of ${\bf 5}$ and $\bar{\bf 5}$ with $B-L$ charges opposite in sign are forbidden by the $B-L$ gauge symmetry.

Assuming that a pair of ${\bf 5}$ and $\bar{\bf 5}$ has an identical $B-L$ charge,%
\footnote{If ${\bf 5}$ and $\bar{\bf 5}$ are embedded in the fundamental representation of $SO(10)$, they can obtain the same charge although an additional $U(1)$ gauge symmetry which commutes with $SO(10)$ is required.}
a minimal way to cancel the gauge anomaly without opposite charges is to introduce five sets of ${\bf 5}$ and $\bar{\bf 5}$~\cite{Nakayama:2011dj}.%
\footnote{From anomaly free conditions of $U(1)_{B-L}^3$, $U(1)_{B-L}-[\text{SM gauge}]^2$, and $U(1)_{B-L}-[{\rm gravity}]^2$, solutions with two (four) pairs of ${\bf 5}$ and $\bar{\bf 5}$ always have one (two) vector-like fermions charged under $U(1)_{B-L}$, $i.e.$ opposite $U(1)_{B-L}$ charges are required. For three pairs of ${\bf 5}$ and $\bar{\bf 5}$, there is no solution due to the Fermat's theorem.}
 One possible charge assignment is 
\begin{align}
\label{eq:c5}
&{\bf 5}(-1),~{\bf 5}(-9),~{\bf 5}(-5),~{\bf 5}(7),~{\bf 5}(8),\\
\label{eq:cb5}
&\bar{\bf 5}(-1),~\bar{\bf 5}(-9),~\bar{\bf 5}(-5),~\bar{\bf 5}(7),~\bar{\bf 5}(8) \ .
\end{align}
As shown in Ref.~\cite{Nakayama:2011dj}, the summation of absolute values of these charge assignments $(-1,-9,-5,~7,~8)$ is the minimum compared to the other charge assignments when the greatest common divisor of the absolute charges is taken as one. The relative normalization of the charges between the above additional sector and the SM sector is not determined by the anomaly free conditions. But, let us assume the charge assignments shown in Eq.\,\eqref{eq:c5} and Eq.\,\eqref{eq:cb5}. Then, some of them can obtain masses by being coupled to $\bar\Phi(10)$,
\begin{align}
W=\bar\Phi(10)\,{\bf 5}(-1)\,\bar{\bf 5}(-9)+\bar\Phi(10)\,{\bf 5}(-9)\,\bar{\bf 5}(-1)+
\bar\Phi(10)\,{\bf 5}(-5)\,\bar{\bf 5}(-5)\ .
\end{align}
Here, we omitted dimensionless couplings for notational simplicity.

To give the other ${\bf 5}$ and $\bar{\bf 5}$ masses, we introduce additional SM gauge singlet chiral superfields with $B-L$ charges $\pm15$,
\begin{align}
\Phi'(-15),~\bar\Phi'(+15)\ ,
\end{align}
which couple to ${\bf 5}$ and $\bar{\bf 5}$ with $+7$ and $+8$ charges,
\begin{align}
W=\Phi'(-15){\bf 5}(8)\bar{\bf 5}(7)+\Phi'(-15){\bf 5}(7)\bar{\bf 5}(8)\ ,
\end{align}
where dimensionless couplings are omitted. The singlets can obtain a VEV by the superpotential,
\begin{align}
W=Y(2\Phi'\bar\Phi'-{v'}^2)\ .
\end{align}
Here, $Y$ is a gauge singlet chiral superfield and $v'$ is a parameter with a mass dimension. 
We also assume that $R$-charges of $Y$, $\Phi'$, and $\bar\Phi$ are $+2$, $0$, and $0$ (see the next paragraph for more details).
In Tab.~\ref{tab:R}, we summarized the field contents and their $B-L$ charges. We note that two $B-L$ breaking fields with different absolute $B-L$ charges like $\Phi$ and $\Phi'$ are needed at least to give all five sets of ${\bf 5}$ and $\bar{\bf 5}$ masses at the renormalizable level.

In addition to the $B-L$ gauge symmetry, we also assume that a discrete subgroup of the $R$-symmetry, ${\mathbb Z}_{NR}~(N>2)$, is a gauge symmetry. 
This assumption is essential to forbid a constant term in superpotential.
However, to produce an almost vanishing cosmological constant after the SUSY breaking, we need a constant term in superpotential which should be generated by a spontaneous breaking of $\mathbb{Z}_{NR}$ to $\mathbb{Z}_{2R}$. Besides this, the $\mu$-term (the Higgsino mass) may be also forbidden by $\mathbb{Z}_{NR}$, and then should be generated by the spontaneous breaking of $\mathbb{Z}_{NR}$ symmetry, which explains the required $\mu$-parameter of order of $1$\,TeV. See Appendix~\ref{app:mu} for more details.
Here, the gauged discrete ${\mathbb Z}_{6R}$ is assumed with the charge assignment in Tab.~\ref{tab:R}, where anomaly free conditions for ${\mathbb Z}_{6R}-SU(3)_c-SU(3)_c$ and ${\mathbb Z}_{6R}-SU(2)_L-SU(2)_L$ are satisfied~\cite{Evans:2011mf}. 
Note that the mixed anomalies of ${\mathbb Z}_{6R}-U(1)_Y-U(1)_Y$ and ${\mathbb Z}_{6R}-U(1)_{B-L}-U(1)_{B-L}$ are model dependent because of the dependence on the normalization of the heavy spectrum~\cite{Krauss:1988zc,Preskill:1990bm,Preskill:1991kd,Banks:1991xj,Ibanez:1991hv,Ibanez:1992ji,Csaki:1997aw,Lee:2010gv,Fallbacher:2011xg,Evans:2011mf}. The gravitational anomaly is also model dependent.%
\footnote{The gravitational anomaly is easily cancelled by introducing some singlet fields under the MSSM and the $B-L$ gauge symmetries.}
 Therefore, we do not specify the field contents from those anomaly cancellations.

\begin{table}[t]
\caption{The charge assignment of the $B-L$ symmetry and the gauged ${\mathbb Z}_{6R}$ symmetry. 
For additional ${\bf 5}$ and $\bar{\bf 5}$, corresponding $B-L$ charges are shown in the parenthesis.
}
\begin{center}
\begin{tabular}{|c||c|c|c|c|c|c|c|c|c|c|c|c|c|}
\hline
 & ${\mathbf{10}}_{\rm SM}$&${\mathbf {\bar 5}}_{\rm SM}$ & $\bar{N}_R$ & $H_u$ & $H_d$  & $\Phi$ &$\bar\Phi$ & $\Phi'$ & $\bar\Phi'$ & $X$ &$Y$ 
  \\ \hline
$U(1)_{B-L}$ & $+1$ &$-3$ &$+5$ &$-2$ &$+2$ & $-10$ & $+10$ & $-15$ & $+15$& $0$ & $0$
\\ \hline
${\mathbb Z}_{6R}$ & $+1/5$ &$-3/5$ &$+1$ &$+8/5$ &$+12/5$ & $0$ & $0$ & $0$& $0$& $+2$& $+2$ 
\\ \hline
\end{tabular}
\end{center}
\begin{center}
\begin{tabular}{|c||c|c|c|c|c|c|c|c|c|c|c|c|c|c|c|}
\hline
 & $\mathbf 5(-1)$&  $\mathbf 5(-5)$&  $\mathbf 5(-9)$&  $\mathbf 5(7)$&  $\mathbf 5(8)$ & $\mathbf{\bar 5}(-1)$ &  $\mathbf{\bar 5}(-5)$ &  $\mathbf{\bar 5}(-9)$ &  $\mathbf{\bar 5}(7)$&  $\mathbf{\bar 5}(8)$  \\ \hline
 ${\mathbb Z}_{6R}$ & $+1$ &$+1$ &$+1$ &$+1$ &$+1$  & $+1$ & $+1$ & $+1$& $+1$& $+1$
\\ \hline
\end{tabular}
\end{center}
\label{tab:R}
\end{table}%

\section{An emergent global Peccei-Quinn symmetry from the B-L gauge symmetry}
\label{sec:PQ}
In regard to the model discussed in Sec.~\ref{sec:model}, we find two accidental global symmetries associated with individual phase rotations of $\Phi$ and $\Phi'$ at the level of the renormalizable Lagrangian. One linear combination of two symmetries corresponds to the $B-L$ symmetry which is gauged in our model. The other combination remains as a global symmetry. 

Let us show the remaining global symmetry is nothing but the PQ symmetry.
The charges of the global symmetry can be chosen so as to satisfy the following generic conditions,
\begin{align}
\label{eq:condition1}
&Q_{\rm MSSM,\bar N_R}=-\frac{q_{\rm B-L}}{10}\times Q_\Phi\ , \\
&Q_{{\bf 5}(-1)}+Q_{\bar{\bf 5}(-9)}=-Q_{\bar\Phi},~Q_{{\bf 5}(-9)}+Q_{\bar{\bf 5}(-1)}=-Q_{\bar\Phi}\ ,\\
&Q_{{\bf 5}(-5)}+Q_{\bar{\bf 5}(-5)}=-Q_{\bar\Phi}\ ,\\
&Q_{{\bf 5}(7)}+Q_{\bar{\bf 5}(8)}=-Q_{\Phi'},~Q_{{\bf 5}(8)}+Q_{\bar{\bf 5}(7)}=-Q_{\Phi'}\ ,\\
&Q_{\Phi}=-Q_{\bar\Phi}\\
&Q_{\Phi'}=-Q_{\bar\Phi'}, \\
\label{eq:condition7}
&Q_{\bar\Phi}/Q_{\Phi'}\neq -2/3\ ,
\end{align}
where $Q_{\rm MSSM,\bar N_R}$ denote the PQ charges of the MSSM fields and right-handed neutrinos, $q_{\rm B-L}$ is the $B-L$ charge, and $Q_X$ denotes  a PQ charge of  a chiral superfield $X$. Charges of the other fields are zero. 
The condition in Eq.\,\eqref{eq:condition7} makes a crucial difference between the global symmetry and $U(1)_{B-L}$.
By using the $B-L$ gauge transformation, the global charge of $Q_\Phi$ can be chosen to be zero.%
\footnote{In other words, zero charges are gauge equivalent to the charge assignments in Eq.\,\eqref{eq:condition1}.}
Then, the above conditions  reduce to 
\begin{align}
\label{eq:condition1_mod}
&Q_\Phi= Q_{\bar\Phi}=0\ ,~Q_{\rm MSSM,\bar N_R}=0\ ,\\
&Q_{{\bf 5}(-1)}+Q_{\bar{\bf 5}(-9)}=Q_{{\bf 5}(-9)}+Q_{\bar{\bf 5}(-1)}=0\ ,~Q_{{\bf 5}(-5)}+Q_{\bar{\bf 5}(-5)}=0\ ,\\
&Q_{{\bf 5}(7)}+Q_{\bar{\bf 5}(8)}=-Q_{\Phi'},~Q_{{\bf 5}(8)}+Q_{\bar{\bf 5}(7)}=-Q_{\Phi'}\ ,\\
\label{eq:condition6_mod}
&Q_{\Phi'}=-Q_{\bar\Phi'}\ .
\end{align}
We have a lot of freedom to choose PQ charges for ${\bf 5}$ and $\bar{\bf 5}$. However, we have additional global $U(1)$ symmetries.
By using those $U(1)$ rotations, we can 
make a choice of PQ charge assignment given in Tab.~\ref{tab:PQ}, where we took $Q_{\Phi'}=-Q_{\bar\Phi'}=+1$, $Q_{{\bf 5}(8)}=Q_{{\bf 5}(7)}=-1$, and the other charges zero.
Now, we see that the global symmetry corresponds to the PQ symmetry, and the extra ${\bf 5}$ and $\bar{\bf 5}$ with $q_{\rm B-L}=7,8$ play a role of the KSVZ quarks~\cite{Kim:1979if,Shifman:1979if}, making the global symmetry anomalous with respect to $SU(3)_{c}$.

\begin{table}[t]
\caption{The charge assignment of accidental PQ symmetry.}
\begin{center}
\begin{tabular}{|c||c|c|c|c|c|c|c|c|c|c|c|c|c|}
\hline
 & ${\mathbf{10}}_{\rm SM}$&${\mathbf {\bar 5}}_{\rm SM}$ & $\bar{N}_R$ & $H_u$ & $H_d$  & $\Phi(-10)$ &$\bar\Phi(+10)$ & $\Phi'(-15)$ & $\bar\Phi'(+15)$ & $X$ &$Y$
 \\ \hline
$PQ$ & $0$ &$0$ &$0$ &$0$ &$0$ & $0$ & $0$ & $+1$ & $-1$ & $0$ & $0$ 
\\ \hline
\end{tabular}
\end{center}
\begin{center}
\begin{tabular}{|c||c|c|c|c|c|c|c|c|c|c|c|c|c|c|c|}
\hline
 & $\mathbf 5(-1)$&  $\mathbf 5(-5)$&  $\mathbf 5(-9)$&  $\mathbf 5(7)$&  $\mathbf 5(8)$ & $\mathbf{\bar 5}(-1)$ &  $\mathbf{\bar 5}(-5)$ &  $\mathbf{\bar 5}(-9)$ &  $\mathbf{\bar 5}(7)$&  $\mathbf{\bar 5}(8)$  \\ \hline
 $PQ$ & $0$ &$0$ &$0$ &$-1$ &$-1$  & $0$ & $0$ & $0$& $0$& $0$
\\ \hline
\end{tabular}
\end{center}
\label{tab:PQ}
\end{table}%

Indeed, under the above PQ charge assignments, we can consider a transformation,
\begin{align}
\label{eq:pqtrans}
\Phi'(-15)&\to e^{i\alpha_{PQ}}\,\Phi'(-15)\ ,\\
{\bf 5}(7)&\to e^{-i\alpha_{PQ}}\,{\bf 5}(7)\ ,\\
{\bf 5}(8)&\to e^{-i\alpha_{PQ}}\,{\bf 5}(8)\ ,
\end{align}
where $\alpha_{PQ}$ denotes the rotation angle.
This rotation leads to the Lagrangian shifts by
\begin{align}
\label{eq:anomaly}
\delta \mathcal{L}_{\cancel{PQ}}=2\alpha_{PQ}\frac{g_s^2}{32\pi^2} G^a_{\mu\nu}\tilde G^{a\mu\nu},
\end{align}
where  $g_s$ is the gauge coupling of $SU(3)_c$, $G^a_{\mu\nu}$ is the gauge field strengths of $SU(3)_c$, and $\tilde G^{a\mu\nu}$ is its dual%
\footnote{$\tilde G^{a\mu\nu}\equiv\frac{1}{2}\epsilon^{\mu\nu\rho\sigma}G^a_{\rho\sigma}$.}.
Note that Eq.\,\eqref{eq:anomaly} seems to show that there is a discrete $\mathbb{Z}_2$ symmetry.
But, this is not correct, $i.e.$ no such a discrete symmetry remains.
This can be seen in Fig.~\ref{fig:interval}, where a domain of the physical interval of the phase component of $\Phi'$ corresponds to $\alpha_{PQ}=[0,\pi)$.%
\footnote{Two points $\alpha_{PQ}=0~\text{and}~\pi$ are equivalent up to the $B-L$ gauge transformation.}
For the domain, there is no degenerate vacuum, and thus there remains no discrete symmetry. We will explain this point more in Sec.~\ref{sec:decomposition}.

The PQ symmetry is explicitly broken by the Planck suppressed operators,
\begin{align}
\bar\Phi(10)^3\Phi'(-15)^2\ ,~\Phi(-10)^3\bar\Phi'(15)^2\ ,
\end{align}
which transform under Eq.\,\eqref{eq:pqtrans} as
\begin{align}
\bar\Phi(10)^3\Phi'(-15)^2\to e^{i2\alpha_{PQ}}\times \bar\Phi(10)^3\Phi'(-15)^2, \\
\Phi(-10)^3\bar\Phi'(15)^2\to e^{-i2\alpha_{PQ}}\times \Phi(-10)^3\bar\Phi'(15)^2\ .
\end{align}
The above operators are forbidden in the superpotential because of the total $R$-charge zero.
Given the fact that the constant term in the superpotential, $W_0$, has the $R$-charge $2$,%
\footnote{The constant $W_0$ can be generated by a VEV of a singlet field. See the Appendix~\ref{app:mu} for more details.}
 we obtain the superpotential,
\begin{align}
\label{eq:wPQbreaking}
W\sim \kappa \frac{W_0}{M_{\rm PL}^3} \frac{\Phi'(-15)^2\bar\Phi(+10)^3}{M_{\rm PL}^2}+\kappa \frac{W_0}{M_{\rm PL}^3} \frac{\bar\Phi'(+15)^2\Phi(-10)^3}{M_{\rm PL}^2}+W_0\ ,
\end{align}
where $\kappa$ is a dimensionless coupling%
\footnote{We took the same coupling $\kappa$ for the first two terms in Eq.\,\eqref{eq:wPQbreaking} for simplicity.}.
The constant term $W_0$ gives the gravitino mass $m_{3/2}=W_0/M_{\rm PL}^2$. Then, the above superpotential terms lead to the scalar potential,
\begin{align}
\label{eq:PQbreakingpotential}
V\sim  \kappa \frac{|m_{3/2}|^2}{M_{\rm PL}} \frac{\phi'(-15)^2\bar\phi(+10)^3}{M_{\rm PL}^2}+\kappa \frac{|m_{3/2}|^2}{M_{\rm PL}} \frac{\bar\phi'(+15)^2\phi(-10)^3}{M_{\rm PL}^2}+c.c.\ ,
\end{align}
where $\phi,~\bar\phi,~\phi'$, and $\bar\phi'$ are the scalar components of the chiral superfields of $\Phi,~\bar\Phi,~\Phi'$, and $\bar\Phi'$, respectively.%
\footnote{The scalar potential is also obtained from the K\"{a}hler potential,
\begin{align} 
K\sim \frac{Z Z^\dagger}{M_{\rm PL}^2}\frac{\phi'(-15)^2\bar\phi(+10)^3}{M_{\rm PL}^2}+\frac{Z Z^\dagger}{M_{\rm PL}^2}\frac{\bar\phi'(15)^2\phi(-10)^3}{M_{\rm PL}^2}+h.c.\ ,
\end{align}
where $Z$ is the SUSY breaking field.}
In the next section, we will see a parameter space where these explicit breaking terms do not spoil the PQ mechanism.

\subsection{Axion and Global PQ}
\label{sec:decomposition}
Before going to the next section, let us decompose the axion and the would-be Nambu Goldstone (NG) boson%
\footnote{The NG boson is eaten by the $B-L$ gauge boson.}
 in the supersymmetric manner.%
\footnote{See Ref.~\cite{Fukuda:2018oco} for more detail discussion about the decomposition in the supersymmetric manner.}
After $\Phi,~\Phi'$ ($\bar\Phi,~\bar\Phi'$) obtain the VEVs, the Goldstone superfields $A_1,~A_2$ are given as%
\footnote{For simplicity, we assume the soft masses of $\Phi$ and $\bar\Phi$ ($\Phi'$ and $\bar\Phi'$) are the same.}
\begin{align}
\label{eq:decomposition1}
&\Phi=\frac{1}{\sqrt{2}}v\,e^{A_1/v}\ ,~\bar\Phi=\frac{1}{\sqrt{2}}v\,e^{-A_1/v}\ ,\\
\label{eq:decomposition2}
&\Phi'=\frac{1}{\sqrt{2}}v'\,e^{A_2/v'}\ ,~\bar\Phi'=\frac{1}{\sqrt{2}}v'\,e^{-A_2/v'}\ .
\end{align}
One linear combination of $A_1$ and $A_2$ corresponds to the would-be NG boson supermultiplet, and the other combination becomes the axion superfield. 
To see this decomposition, let us consider the K\"{a}hler potential,
\begin{align}
K=\Phi^\dagger e^{-2\times (-10)\times gV}\Phi+\bar\Phi^\dagger e^{-2\times (+10)\times gV}\bar\Phi+{\Phi'}^\dagger e^{-2\times (-15)\times gV}{\Phi'}+\bar\Phi'^\dagger e^{-2\times (+15)\times gV}\bar\Phi'
\label{eq:Kahler}
\end{align}
where $V$ and $g$ denote the $B-L$ gauge supermultiplet and its gauge coupling constant. As a result of substituting Eq.\,\eqref{eq:decomposition1} and Eq.\,\eqref{eq:decomposition2} into Eq.~(\ref{eq:Kahler}), the K\"{a}hler potential becomes
\begin{align}
K=v^2\cosh\left(2\times (-10)gV-\frac{A_1+A_1^\dagger}{v}\right)+v'^2\cosh\left(2\times (-15)gV-\frac{A_2+A_2^\dagger}{v'}\right)\ .
\end{align}
Then, the axion and would-be NG boson superfields $A$ and $G$ are defined as
\begin{align}
\left(
\begin{array}{c}
A^{(\dagger)}\\
G^{(\dagger)}
\end{array}
\right)
=
\frac{1}{\sqrt{(10)^2v^2+(15)^2v'^2}}
\left(
\begin{array}{cc}
15v' & -10v\\
 10v & 15v'
\end{array}
\right)
\left(
\begin{array}{c}
A_1^{(\dagger)}\\
A_2^{(\dagger)}
\end{array}
\right)\ ,
\end{align}
and then the K\"{a}hler potential can be rewritten as
\begin{align}
K=v^2\cosh\left(2\times (10)g\tilde V+g\frac{2\times15}{m_V}\frac{v'}{v}(A+A^\dagger)\right)+v'^2\cosh\left(2\times (15)g\tilde V-g\frac{2\times10}{m_V}\frac{v}{v'}(A+A^\dagger)\right)\ .
\end{align}
Here, 
\begin{align}
&\tilde V\equiv V+\frac{1}{m_V}(G+G^\dagger)\ ,\\
&m_V \equiv 2g\sqrt{(10)^2v^2+(15)^2v'^2}\ .
\end{align}
Note that the axion $A$ is invariant under the $B-L$ gauge transformation.

\begin{figure}[ht]
\begin{center}
  \includegraphics[width=.7\linewidth]{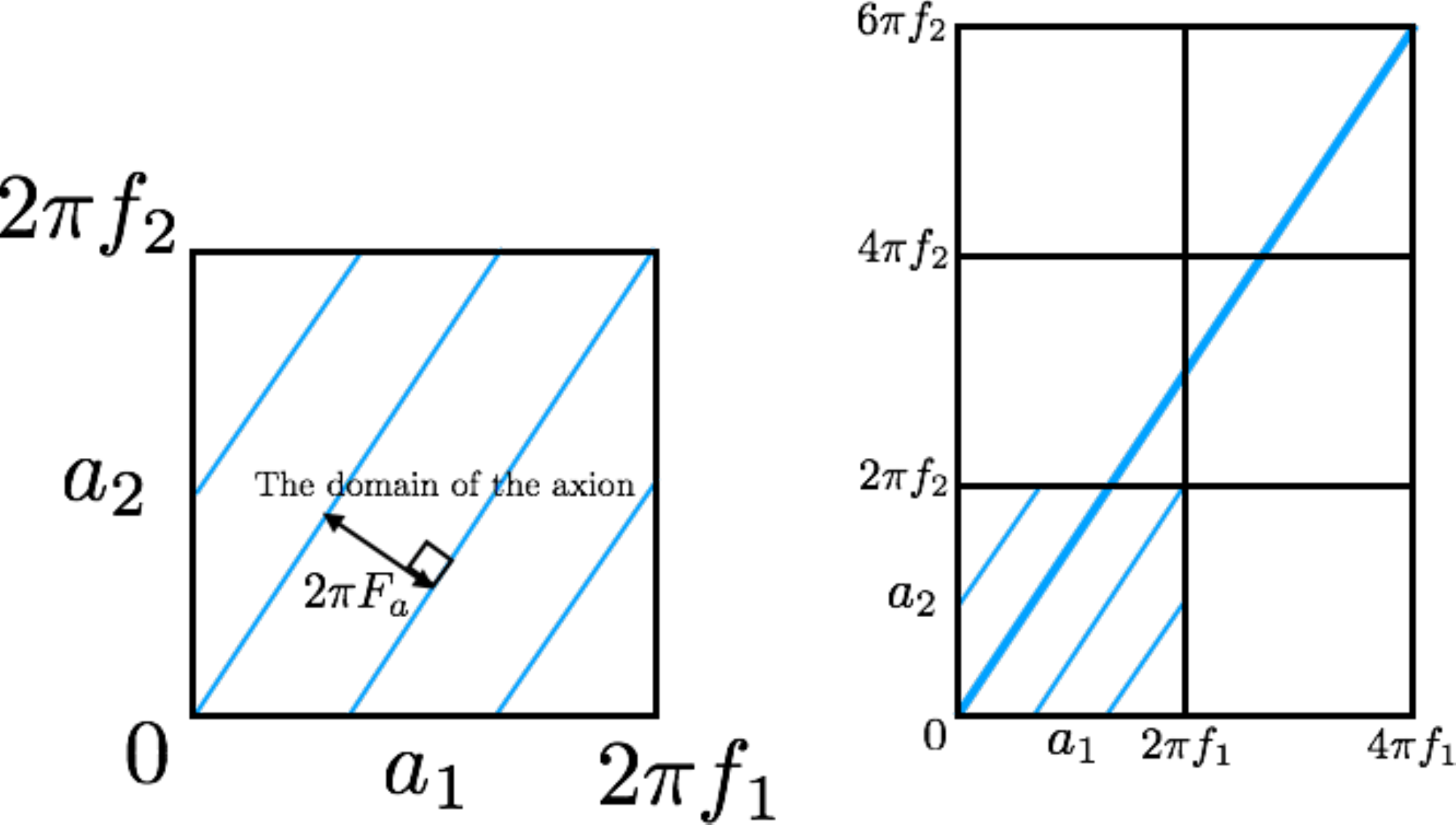}
 \end{center}
\caption{\sl \small (Left) A gauge orbit in the domain of  $(a_1,a_2)$ for $|q|=10,\,|q'|=15$. (Right) The  unwound gauge orbits.
}
\label{fig:interval}
\end{figure}

Let us discuss the domain and the effective decay constant of the axion. The domains of the phases of $\Phi$ and $\Phi'$ are given as
\begin{align}
\frac{a_1}{f_1}\equiv\frac{{\rm Im}[\tilde A_1]}{v}=[0,2\pi)\ ,\\
\frac{a_2}{f_2}\equiv\frac{{\rm Im}[\tilde A_2]}{v'}=[0,2\pi)\ .
\end{align}
Here, $f_1=\sqrt{2}v$, $f_2=\sqrt{2}v'$, $\tilde A_i$ denotes the scalar component of $A_i$, and $a_i=\sqrt{2}{\rm Im}[\tilde A_i]$. The domain of the axion is obtained as the interval of the gauge orbit~\cite{Fukuda:2017ylt} and we find the domain from Fig.~\ref{fig:interval},
\begin{align}
a\equiv\sqrt{2}\,{\rm Im}[\tilde A]=[0,2\pi\,F_a)\ ,
\end{align}
where $\tilde A$ denotes the scalar component of $A$, $a=\sqrt{2}{\rm Im}[\tilde A]$ denotes the axion, and $F_a$ is given by 
\begin{align}
\label{eq:axiondecayconstant}
F_a=\frac{\sqrt{2}vv'}{\sqrt{4v^2+9v'^2}}\ ,
\end{align}
as an effective axion decay constant. 
We note that Eq.\,\eqref{eq:axiondecayconstant} is  valid under the assumption of $\langle\Phi\rangle=\langle\bar\Phi\rangle$ and $\langle\Phi'\rangle=\langle\bar\Phi'\rangle$.
If this assumption about the VEVs does not hold, the kinetic term of the axion is not canonically normalized, and thus the replacement of $v$ ($v'$) into $\sqrt{\langle\Phi\rangle^2+\langle\bar\Phi\rangle^2}$ ($\sqrt{\langle\Phi'\rangle^2+\langle\bar\Phi'\rangle^2}$) is needed in Eq.\,\eqref{eq:axiondecayconstant}. 

After extra fields of ${\bf 5}$ and $\bar{\bf 5}$ are integrated out, the axion has a coupling~\cite{Fukuda:2017ylt},
\begin{align}
\mathcal{L}=\frac{a}{F_a}\frac{g_s^2}{32\pi^2} G^a_{\mu\nu}\tilde G^{a\mu\nu}\ .
\end{align}
This shows that there remains no discrete symmetry, $i.e.$ the domain wall number is one ($N_{\rm dw}=1$).

\subsection{Domain Wall Problem}
It should be remarked that the model may suffer from the domain wall problem~\cite{Fukuda:2018oco,Hiramatsu:2019tua}.
For example, let us consider a case where the phase transition of $\Phi\bar\Phi\neq 0$ occurs before/during inflation while the transition of $\Phi'\bar\Phi'\neq 0$ takes place after the inflation ends.%
\footnote{For simplicity, we assume $\Phi\simeq\bar\Phi$ and $\Phi'\simeq\bar\Phi'$ after the phase transitions}
In the first transition of $\Phi\bar\Phi\neq 0$, $U(1)_{B-L}$ symmetry is spontaneously broken, and thus cosmic strings are formed while they are inflated away. In the second transition of $\Phi'\bar\Phi'\neq 0$, $U(1)_{PQ}$ symmetry is spontaneously broken, and the comic strings are formed. Let us call this string $\Phi'$-string.
Around the $\Phi'$-string with the winding number one, the phase of $a_2/f_2$ changes from $0$ to $2\pi$ because the gauge freedom is effectively frozen, since the local strings are inflated away. In this domain, the axion potential gives rise to the energy contrast and the axion distribution crosses the potential minimum two times (see also Fig.~\ref{fig:interval}). This shows the $\Phi'$-string is attached by two domain walls. Therefore, this case corresponds to the domain wall number two scenario effectively. Similarly, when $\Phi'\bar\Phi'\neq0$ takes place during the inflation while $\Phi\bar\Phi\neq0$ does after the inflation ends, the comic strings attached by three domain walls will be formed ($\Phi$-string).
In the case where both phase transitions occur after inflation, the cosmic string network will become much more complicated due to the coexistence of $\Phi$ and $\Phi'$-strings. We need detailed numerical analysis in this case, which is beyond the scope of this paper.

Therefore, in this paper, we focus on the case where both phase transitions take place before/during inflation. Then, every defect is inflated away and domain wall problem can be avoided.%
\footnote{Even if both phase transitions take place before/during inflation, the domain wall problem may not be avoided when the field values are large for example $\Phi\sim \bar\Phi' \sim M_{\rm PL}$ during inflation because the axion fluctuations are produced by the parametric resonance after inflation~\cite{Kasuya:1996ns}.}

\section{A consistent axion model with quantum gravity effects and its low energy phenomenology}
\label{sec:pheno}
In the presence of the explicit PQ breaking terms in Eq.\,\eqref{eq:PQbreakingpotential}, the QCD vacuum angle is shifted as
\begin{align}
\label{eq:dtheta}
{\mit \Delta}\theta\simeq \kappa\frac{m_{3/2}^2\langle\phi(-10)\rangle^3\langle \bar\phi'(+15)\rangle^2}{M_{\rm PL}^3m_a^2F_a^2}\simeq 10^{-9}\kappa\left(\frac{m_{3/2}}{1\,{\rm eV}}\right)^2\left(\frac{\langle \phi\rangle}{10^{12}\,{\rm GeV}}\right)^3\left(\frac{\langle\phi'\rangle}{10^{12}\,{\rm GeV}}\right)^2\ ,
\end{align}
where $m_a$ denotes the axion mass, and $\langle \phi\rangle\simeq\langle \bar\phi\rangle$ and $\langle \phi'\rangle\simeq\langle \bar\phi'\rangle$ are assumed.
This small shift should satisfy the condition
\begin{align}
\label{eq:thetabound}
{\mit \Delta}\theta\lesssim 10^{-10}
\end{align}
to be consistent with the experimental bound on the $\theta$ angle~\cite{Baker:2006ts}. We note that we have not found any explicit PQ breaking terms which have a smaller Planck mass suppression and lead to  a larger shift of the angle compared to Eq.\,\eqref{eq:dtheta}.%

Another constraint on the model is derived from the domain wall problem. 
As we have already mentioned in the previous section,
the model suffers from the domain wall problem because it is expected that there appear strings with the domain wall number larger than one on $B-L$ and PQ symmetry breaking~\cite{Fukuda:2018oco}. To avoid this problem, 
we consider the case where $\Phi$ ($\Phi'$) obtains a non-zero field value during inflation.
We also assume a similar size of positive Hubble induced masses for $\Phi$ and $\bar\Phi$ ($\Phi'$ and $\bar\Phi'$), and then they obtain a field value around $v$ ($v'$) during inflation.%
\footnote{The axion has the coupling with the inflaton field $I$, $
    K\sim \frac{|I^2|\Phi^3\bar\Phi'^2}{M_{\rm PL}^5}+\frac{|I^2|\bar\Phi^3\Phi'^2}{M_{\rm PL}^5}$.
This leads to the Hubble induced mass for the axion as large as $H_{\rm inf}$ if $\Phi$ and $\bar\Phi'$ ($\Phi'$ and $\bar\Phi$) obtain the field value around the Planck scale during inflation. But, once the fields start to oscillate after the inflation end, the fluctuations of the axion are produced through the parametric resonance, and then the domain wall problem may be formed~\cite{Kasuya:1996ns} as we have already mentioned. Thus, we focus on the positive Hubble induced mass case in this paper.
See also Ref.\,\cite{Kawasaki:2017kkr} for more details.}
Once the PQ symmetry is broken during inflation, the axion  develops fluctuations because the axion is almost massless at the moment.
After the reheating, below the temperature of the QCD scale, the axion acquires its mass and starts coherent oscillation.
Then, the axion field fluctuations turn into the isocurvature density fluctuations of the axion and the power spectrum thereof $\mathcal{P}_{\rm ISO}$ is given by (see $e.g.$ Ref.~\cite{Kawasaki:2013ae})
\begin{align}
\mathcal{P}_{\rm ISO}\simeq 6.8\left(\frac{H_{\rm inf}}{2\pi\,F_a}\right)^2\left(\frac{F_a}{10^{12}\,{\rm GeV}}\right)^{1.19}\left(\frac{\Omega_ah^2}{0.12}\right)\ ,
\label{eq:Piso}
\end{align}
Here, $\Omega_ah^2$ denotes the axion abundance,
\begin{align}
\Omega_ah^2\simeq 0.18\theta_a^2\left(\frac{F_a}{10^{12}\,{\rm GeV}}\right)^{1.19}\ ,
\end{align}
where $\theta_a$ is an initial misalignment angle~\cite{Lyth:1991ub}. From CMB observations, the isocurvature perturbations at the pivot scale $k_{*}\simeq0.05{\rm Mpc}^{-1}$ are constrained to be~\cite{Nunez:2018zrh}
\begin{align}
\frac{\mathcal{P}_{\rm ISO}}{\mathcal{P}_\zeta+\mathcal{P}_{\rm ISO}}\leq 0.038\ ,
\end{align}
where $\mathcal{P}_\zeta(\simeq2.2\times10^{-9})$~\cite{Ade:2015xua} is the power spectrum of the curvature perturbations. Then, we obtain 
\begin{align}
\label{eq:hinf}
H_{\rm inf}\lesssim 1.8\times 10^7\,{\rm GeV} \left(\frac{F_a}{10^{12}\,{\rm GeV}}\right)^{-0.19}
\end{align}
where the initial misalignment angle is taken as $\mathcal{O}(1)$.%

In addition, to avoid the restoration of the two symmetries, we require the condition for the maximum temperature $T_{\rm max}$ during reheating~\cite{Kolb:1990vq,Chung:1998rq}
\begin{align}
\label{eq:tmax}
T_{\rm max}\simeq 0.5\times T_R^{1/2}H_{\rm inf}^{1/4}M_{\rm PL}^{1/4}\lesssim {\rm min}[\langle\Phi\rangle,~\langle\Phi'\rangle] \,
\end{align}
where $T_R$ denotes a reheating temperature and we take the effective massless degrees of freedom to be about $\sim200$.%
\footnote{We are assuming that SUSY particles are also 
in the thermal bath.}
Let us also consider the cosmology of the extra ${\bf 5}$ and $\bar{\bf 5}$. They will obtain masses around $\langle\Phi\rangle$ or $\langle\Phi'\rangle$ and they are stable because of $U(1)_{B-L}$ and $Z_{6R}$.%
\footnote{
We have not found any processes which lead to the entire decays of the extra ${\bf 5}$ and $\bar{\bf 5}$ into the MSSM particles.
}
Therefore, once they are produced in the thermal plasma, they will lead to the overclosure of the universe. To avoid this problem, we require the condition for the heavy particle mass and  $T_{\rm max}$~\cite{Chung:1998rq},
\begin{align}
\label{eq:tmax2}
    T_{\rm max}\lesssim \frac{1}{10} \times{\rm min}[\langle\Phi\rangle,~\langle\Phi'\rangle]\ .
\end{align}

Let us discuss the gravitino mass bound. From Eq.\,\eqref{eq:dtheta} and Eq.\,\eqref{eq:thetabound}, the gravitino mass is upper-bounded for given values of the VEVs of scalars. Using the lower bound on the axion decay constant $F_a\gtrsim 10^8$\,GeV~%
\footnote{$F_a\gtrsim 10^8$\,GeV is consistent with the constraint from the supernova 1981A observation~\cite{Chang:2018rso}.}
and the requirement of ${\mit \Delta}\theta \lesssim 10^{-10}$, we obtain the gravitino mass upper-bound,%
\footnote{Recently, there is a debate on the supernova cooling bound~\cite{Bar:2019ifz}. Neglecting the supernova constraint and using $F_a\gtrsim 10^7$\,GeV~\cite{Cadamuro:2011fd} open up new possibilities of $m_{3/2}\lesssim 0.1-1\,{\rm TeV}$.}
\begin{align}
m_{3/2}\lesssim \frac{1}{\sqrt{\kappa}}1\,{\rm GeV}\ .
\label{eq:m32ub}
\end{align}
We consider the gauge mediation model to obtain the light gravitino mass (see the following discussion for more details).

For gravitino cosmology, as long as the reheating temperature $T_R$ is higher than SUSY particle masses, a light gravitino of mass $\lesssim1$\, GeV is produced through scattering processes of MSSM or messenger particles and its abundance exceeds easily the DM density if $m_{3/2}\gtrsim100$\,eV.%
\footnote{The gravitino number density can be diluted if we have an enough entropy production at a later time. However, we concentrate our discussion  on the case in the absence of such a late time entropy production, for simplicity, in this paper.}

Thus, we consider
the gravtino mass $m_{3/2}\lesssim 100$\,eV.
However, this mass range is already excluded by the observation of Ly-$\alpha$ forests~\cite{Viel:2005qj} because the gravitino behaves as warm dark matter. Besides, by the use of recent data from the observations of the CMB lensing and the cosmic shear, the gravitino mass is upper-bounded by $m_{3/2}\lesssim 4.7\,{\rm eV}$~\cite{Osato:2016ixc}. 
This mass range is compatible with many baryogenesis scenarios of high reheating temperature typified by the thermal leptogenesis. In the following discussion, we concentrate on the parameter space with $m_{3/2}\simeq \mathcal{O}(1)\,{\rm eV}$. 

\begin{figure}[ht]
\begin{center}
\begin{minipage}{\linewidth}
  \includegraphics[width=.4\linewidth]{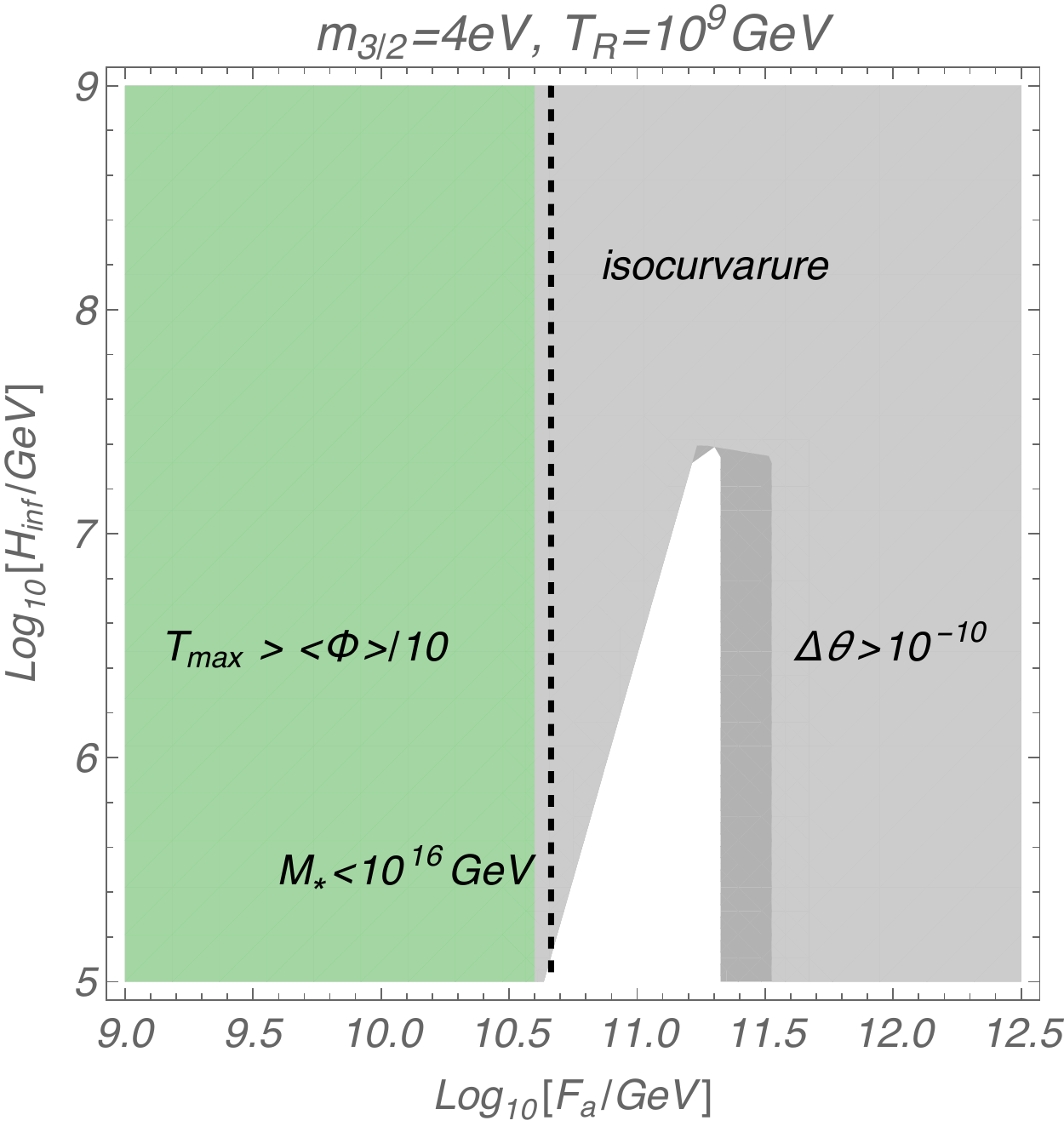}
\begin{minipage}{0.06\hsize}
        \hspace{2mm}
      \end{minipage}  
  \includegraphics[width=.4\linewidth]{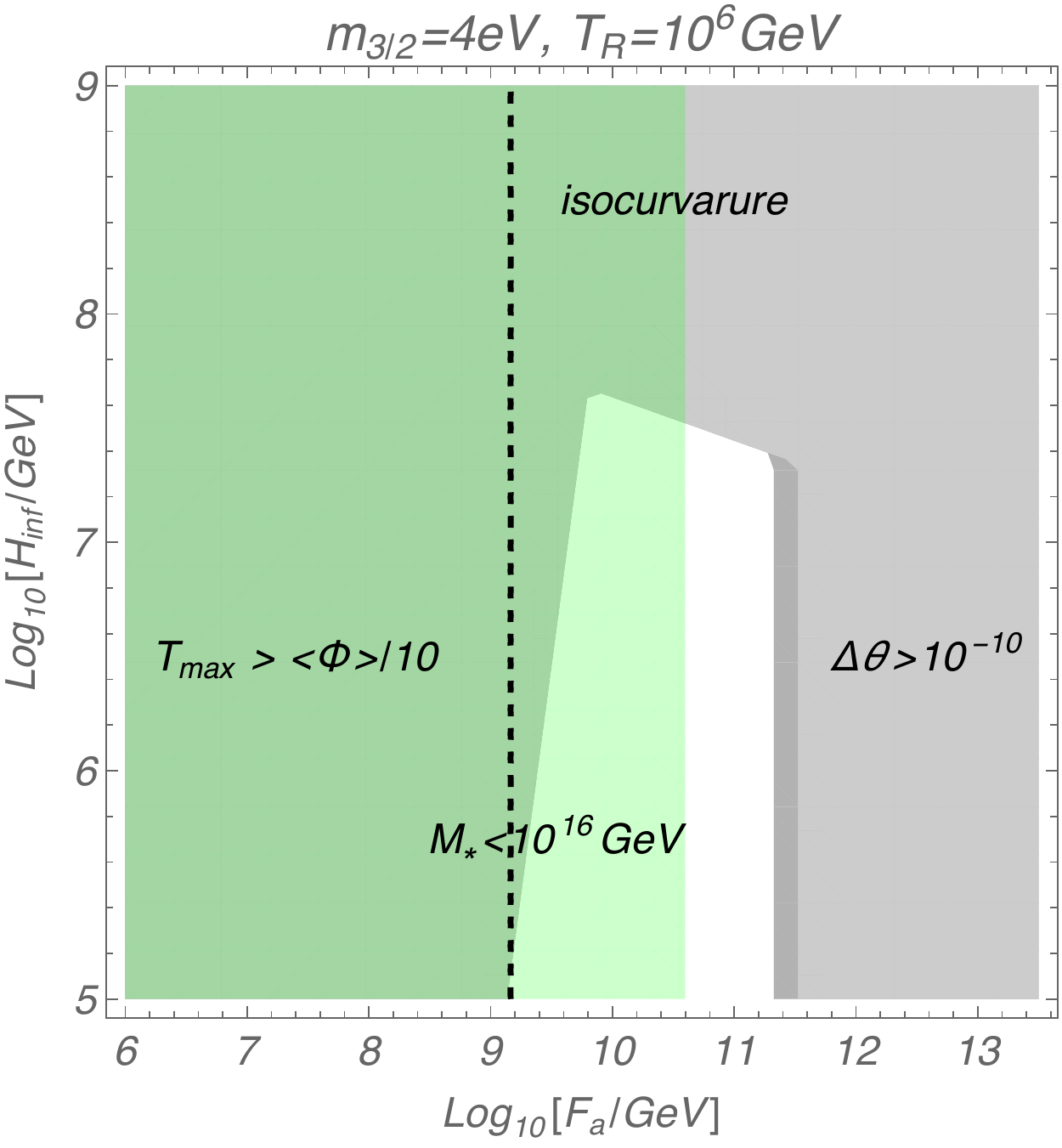}
\end{minipage}
 \end{center}
\caption{\sl \small 
(Left) Constraints on $F_a-H_{\rm inf}$ for $m_{3/2}=4\,{\rm eV}$ and $T_R=10^9\,{\rm GeV}$. (Right) Constraints on $F_a-H_{\rm inf}$ for $m_{3/2}=4\,{\rm eV}$ and $T_R=10^6\,{\rm GeV}$. The darker (lighter) gray shaded region corresponds to the constraint of $\kappa=1~(10^{-1})$ in Eq.\,\eqref{eq:dtheta}. In the green shaded regions, the gauge coupling constants become non-perturbative at a scale below $10^{16}$\,GeV. On the black dashed line, the current dark matter abundance is explained by the non-thermally produced saxion and axino. (For smaller $F_a$, the abundance is larger.) See more details in the main text.
}
\label{fig:constraint1}
\end{figure}

In Fig.~\ref{fig:constraint1}, we show the allowed parameter space from the above constraints. The gray shaded region is constrained from three conditions in Eq.\,\eqref{eq:thetabound},~Eq.\,\eqref{eq:hinf},~Eq.\,\eqref{eq:tmax},~and Eq.\,\eqref{eq:tmax2}.
The darker (lighter) gray shaded region corresponds to the case of $\kappa=1~(0.1)$ in Eq.\,\eqref{eq:dtheta}.
 In the figure, the gravitino mass is taken as $m_{3/2}=4\,{\rm eV}$  and $\langle\Phi\rangle=\langle\Phi'\rangle$ is assumed. We are also taking the reheating temperature $T_R=10^9\,{\rm GeV}~(T_R=10^6\,{\rm GeV})$ for the left (right) figure. 
The axion becomes the dominant dark matter component for about $F_a\simeq 10^{11}$\,GeV assuming an initial misalignment angle around $\pi$~(see $e.g.$ Ref.~\cite{Kawasaki:2013ae}).

The gravitino mass around $1$\,eV is provided by low-scale gauge mediation models (see $e.g.$ Refs.~\cite{izawa:1997gs,Csaki:2006wi,Dine:2006xt,Shirai:2010rr,Ibe:2010jb,Evans:2011pz,Ibe:2016kyg}).
In the weakly coupled low scale gauge mediation models, however, the large gravitino mass $m_{3/2}\gtrsim 10$\,eV is required to explain the observed Higgs boson mass by heavy SUSY particles with masses around $\mathcal{O}(10)$\,TeV~\cite{Ajaib:2012vc,Yanagida:2012ef}. The gravitino mass can be as light as $\mathcal{O}(1)$\,eV in the strongly coupled low scale gauge mediation models~\cite{Ibe:2010jb}.  Indeed, the Higgs boson mass is explained by $e.g.$ $N_{\rm mess}=4$, $M_{\rm mess}\simeq 10^5$\,GeV, and $F^{1/2}_{m}\simeq10^5$\,GeV~\cite{Ibe:2016kyg}. Here, $N_{\rm mess}$ denotes a number of pairs of the chiral superfields (messengers) in the fundamental and anti-fundamental representations of $SU(5)$, $M_{\rm mess}$ is the messenger mass scale, and  $F^{1/2}_{m}$ is the mass splitting between the messenger scalars and the fermions.

Including $N_{\rm mess}$ pairs of ${\bf 5}$ and $\bar{\bf 5}$ as the messengers in the gauge mediation model, $N_{\rm mess}+5$ pairs of extra multiplets make the renormalization group running of the gauge coupling constants non-asymptotically free. In the green shaded region in Fig.~\ref{fig:constraint1}, at least one of the MSSM gauge coupling constants becomes larger than $4\pi$ at a scale below  $10^{16}$\,GeV. Here, we use the one-loop renormalization group assuming  $N_{\rm mess}=4$, all messengers with $10^5$\,GeV mass, and all MSSM SUSY particles with $1$\,TeV.%
\footnote{The contributions of the messengers to the running of the MSSM gauge coupling constants can be reduced if a hidden gauge theory of the messengers is embedded in a conformal field theory at high energies and the anomalous dimensions of messengers are rendered positive by a large Yukawa coupling between the messengers and some hidden matters~\cite{Sato:2009yt}.}
 We also take the mass of the other five pairs of ${\bf 5}$ and $\bar{\bf 5}$ as $\langle\Phi\rangle=\langle\Phi'\rangle$.

Let us also discuss cosmology of the saxion and the axino. Both saxion and axino can obtain masses around $m_{3/2}\simeq 1\,{\rm eV}$ 
and they are stable with lifetime larger than the age of the Universe~(see $e.g.$ Ref.~\cite{Kawasaki:2013ae}).
These light particles are non-thermally produced by $e.g.$ the gluino and gluon scatterings. The abundance of saxion and axino is~\cite{Covi:2001nw,Brandenburg:2004du,Strumia:2010aa,Graf:2012hb},
\begin{align}
\Omega_{\text{non-th}} h^2\simeq 6\times 10^{-3}g_s(T_R)^6\left(\frac{m_{3/2}}{1\,{\rm eV}}\right)\left(\frac{F_a}{10^{11}\,{\rm GeV}}\right)^{-2}\left(\frac{T_R}{10^9\,{\rm GeV}}\right)\ ,
\end{align}
where $g_s(T_R)$ is the $SU(3)_c$ gauge coupling at the energy scale $T_R$.
In Fig.~\ref{fig:constraint1}, on the black dashed line, the current dark matter abundance is explained by saxion and axino.%
\footnote{We took $g_s(T_R)=0.9$.}
 On the left-hand region of the dashed line, the universe is over-closed.
 Besides, for the saxion, its coherent oscillation also contributes to the abundance~\cite{Kim:1992eu,Hashimoto:1998ua,Asaka:1998xa,Banks:2002sd},
\begin{align}
\Omega_{\text{osc}} h^2\simeq 2\times 10^{-3} \left(\frac{m_{3/2}}{1\,{\rm eV}}\right)^{1/2}\left(\frac{F_a}{10^{11}\,{\rm GeV}}\right)^2\left(\frac{\sigma_I}{F_a}\right)^2\ ,
\end{align}
where $\sigma_I$ is an initial amplitude of the oscillation. Assuming a positive Hubble induced mass for the saxion, the amplitude will be as large as $\sigma_I\simeq H_{\rm inf}\lesssim F_a$.
Then, the saxion abundance from the saxion oscillation is estimated to be much smaller than the current dark matter abundance.

\section{Conclusions}
\label{sec:conclusion}
In this paper, we argue that an origin of an accidental PQ symmetry with good quality can be provided by a simple extension of the MSSM with $B-L$ gauge symmetry. The model is based on the MSSM and three right-handed neutrinos. We introduced five pairs of ${\bf 5},~\bar{\bf 5}$ and some MSSM gauge singlet fields which are charged under the $B-L$ gauge symmetry. In Sec.~\ref{sec:PQ}, we show that the model can enjoy an accidental PQ symmetry with the help of the $B-L$ gauge symmetry. On top of this, the non anomalous gauged $\mathbb{Z}_{6R}$ symmetry in the model helps improving the axion quality. The explicit PQ breaking operators are suppressed by the $B-L$ gauge symmetry and a discrete gauged $R$-symmetry $\mathbb{Z}_{6R}$. 
We showed the axion quality problem can be solved by the small gravitino mass $m_{3/2}\ll 100$\,GeV. An interesting parameter range of the gravitino mass is $m_{3/2}=\mathcal{O}(1)$\,eV which is consistent with many baryogenesis scenarios enabled by the thermal leptogenesis. In that range, we found a viable parameter space of the axion decay constant $F_a\simeq 10^{11}$\,GeV and $H_{\rm inf}\ll 10^{12}$\,GeV.
Let us comment on the testability of the model. If the whole DM is attributed to the light gravitino, its free-streaming would affect the large scale structure in the universe as warm dark matter, and thus it is severely constrained from astrophysical and cosmological observations as we mentioned above. Nevertheless, the presence of such a light gravitino can be still allowed provided it contributes to the current DM abundance only at a partial level. For this case, interestingly, the gravitino dark matter even with $m_{3/2}\simeq 1$\,eV may be tested by future observation of the $21$ cm line fluctuations~\cite{Oyama:2016lor}.

Along with the capability of the model to address the strong CP problem by having an emergent PQ symmetry, we emphasize a natural generation of the $\mu$-term (Higgsino mass term) as an another virtue of the model. The axion solution to the strong CP problem demanding $m_{3/2}\lesssim1\,{\rm GeV}$ in the model (see Eq.~(\ref{eq:m32ub})), taking $\mathbb{Z}_{6R}$ as the discrete $R$-symmetry was a rather unartificial option. As an accompanying result, the model was shown to be able to naturally produce the $\mu$-term of the right size around TeV scale without small parameters, relying on the spontaneous breaking of $\mathbb{Z}_{6R}$ to $\mathbb{Z}_{2R}$ discussed in Appendix~\ref{app:mu}.

Finally, let us comment that the constraint on the gravitino mass can be relaxed if we have an extra dimension.
The explicit PQ breaking operators in Eq.\,\eqref{eq:PQbreakingpotential} can be more suppressed in an extra-dimensional setup~\cite{Cheng:2001ys,Izawa:2002qk}, where the MSSM, right-handed neutrinos, $\Phi(\bar\Phi)$, and ${\bf 5}$ and $\bar{\bf 5}$ with $-1,-5,-9$ charges of $U(1)_{B-L}$ reside on a brane while $\Phi'(\bar\Phi')$ and ${\bf 5}$ and $\bar{\bf 5}$ with $7, 8$ charges of $U(1)_{B-L}$ sit on a separated brane. If a distance between two branes is larger than $70\,M_{5}^{-1}$, we may have $m_{3/2} >100$\,TeV, where we consider a (4+1) dimensional space time and $M_5$ is a cutoff scale in the theory. Or we can take the $B-L$ breaking scale~$10^{15}$ GeV keeping $m_{3/2}=\mathcal{O}(1)$\,eV if the distance is larger than $30\,M_{5}^{-1}$. If it is the case, we may have a new DM candidate found in~\cite{Choi:2020nan} in a framework of the $B-L$ gauge symmetry.

\vspace{-.4cm}  
\begin{acknowledgments}
\vspace{-.3cm}
T. T. Y. is supported in part by the China Grant for Talent Scientific Start-Up Project and the JSPS Grant-in-Aid for Scientific Research No. 16H02176, No. 17H02878, and No. 19H05810 and by World Premier International Research Center Initiative (WPI Initiative), MEXT, Japan.
M. S. and T. T. Y. thank Kavli IPMU for their hospitality during the corona virus pandemic.
\end{acknowledgments}

\appendix
\section{Generating constant term in the superpotential and $\mu$-term}
\label{app:mu}
To generate a constant term in the superpotential, we introduce a chiral superfield $S$ with $R$-charge $+2$. The superpotential is given by
\begin{align}
W=\Lambda^2S+\frac{\lambda}{M_{\rm PL}} S^4\ ,
\end{align}
which is allowed by $\mathbb{Z}_{6R}$. Here, $\Lambda$ is a parameter with a mass dimension%
\footnote{The $\Lambda^2S$ term can be generated by the strong dynamics of the hidden $SU(2)$ (see the following discussion).} and $\lambda$ is a dimensionless coupling. 
$S$ obtains the non-zero VEV,
\begin{align}
\langle S\rangle = \left(\frac{\Lambda^2M_{\rm PL}}{4\lambda}\right)^{1/3}\ ,
\end{align}
satisfying the F-term condition for $S$. 
Suppose 
\begin{align}
\label{eq:param1}
    \lambda=\mathcal{O}(1),~\Lambda\simeq 10^{8}\,{\rm GeV}\ ,
\end{align}
and then we obtain 
\begin{align}
\label{eq:svev}
    \langle S\rangle\simeq 10^{11}\,{\rm GeV}\ ,
\end{align} 
and this yields the VEV of the superpotential,
\begin{align}
W_0\simeq 5.5\times 10^{27}\,{\rm GeV}^3\simeq m_{3/2}M_{\rm PL}^2\ ,
\end{align}
for $m_{3/2}\simeq 1$\,eV. The following coupling between $S$ and $H_uH_d$ is allowed in the superpotential,
\begin{align}
\label{eq:mu}
W\sim \frac{S^2}{M_{\rm PL}}H_uH_d\ ,
\end{align}
which can lead to the Higgsino mass around $\mathcal{O}(1)$\,TeV ($\mu$-term) on the spontaneous breaking of $Z_{6R}$.

While $W_0$ and $\mu$-term are generated in the above setup, we meet a problem of large explicit PQ breaking terms. Indeed, $S$ couples to the PQ breaking operators,
\begin{align}
W\sim \frac{S}{M_{\rm PL}}\frac{\Phi^3\bar\Phi'^2}{M_{\rm PL}^2}+\frac{S}{M_{\rm PL}}\frac{\bar\Phi^3\Phi'^2}{M_{\rm PL}^2}\ .
\end{align}
Then, the scalar potential,
\begin{align}
V\sim  \frac{m_{3/2}^*\langle S\rangle}{M_{\rm PL}} \frac{\phi'(-15)^2\bar\phi(+10)^3}{M_{\rm PL}^2}+\frac{m_{3/2}^*\langle S\rangle}{M_{\rm PL}} \frac{\bar\phi'(+15)^2\phi(-10)^3}{M_{\rm PL}^2}+c.c.\ ,
\end{align}
is obtained, which can make a dangerous contribution to the $\theta$ angle. To solve the quality problem, the gravitino mass must be much smaller than $1$\,eV which cannot be obtained even in low scale gauge mediation models. 

The above problematic potential can be suppressed by considering an additional $\mathbb{Z}_4$ discrete gauge symmetry. Let us consider that $\Lambda$ is a spectator field%
\footnote{We can consider that $\Lambda$ is given by a strong dynamics. For example, let us consider the hidden $SU(2)$ gauge theory with four fundamental representation chiral superfields $Q_i~(i=1-4)$ with $+1$ charge under $\mathbb{Z}_4$ and zero charge under $\mathbb{Z}_{6R}$. Below the dynamical scale of $\tilde\Lambda$, the model is described by the composite states of mesons. The mesons may obtain the VEV from the quantum modified constraint. Then, the term $W=\Lambda^2S$ is obtained from $W=\lambda_SSQQ\simeq \lambda_S \tilde \Lambda^2S$ where $\lambda_S$ is a dimensionless coupling and flavor indices are omitted for simplicity. }
 with the charge $+1$ under new $\mathbb{Z}_4$ symmetry%
\footnote{The $R$-charge of $\Lambda$ is taken as zero.}
 and $S$ also has charge $+2$ under $\mathbb{Z}_4$. The Higgs fields $H_u,~H_d$ have zero charges of  $\mathbb{Z}_4$, and then $\mu$-term in Eq.~\eqref{eq:mu} is allowed. On the other hand, the explicit PQ breaking terms are suppressed by
\begin{align}
W\sim \frac{\Lambda^2S}{M_{\rm PL}^3}\frac{\Phi^3\bar\Phi'^2}{M_{\rm PL}^2}+\frac{\Lambda^2S}{M_{\rm PL}^3}\frac{\bar\Phi^3\Phi'^2}{M_{\rm PL}^2}+ \frac{S^4}{M_{\rm PL}^4}\frac{\Phi^3\bar\Phi'^2}{M_{\rm PL}^2}+\frac{S^4}{M_{\rm PL}^4}\frac{\bar\Phi^3\Phi'^2}{M_{\rm PL}^2}\ .
\end{align}
The order of the explicit PQ breaking is the same as Eq.\,\eqref{eq:wPQbreaking}, and thus the suppression is enough. 

Finally, let us comment on the domain wall problem related to $S$.
Let us first consider the case where $S$ obtains the positive Hubble induced mass and sits around the origin during the inflation. After the end of the inflation, it rolls down to the potential minimum of $S=\langle S\rangle$, and then the domain walls are formed by the discrete symmetry breaking of $\mathbb{Z}_{6R}$ into $\mathbb{Z}_{2R}$. On the other hand, for the negative Hubble induced mass, $S$ obtains the non-zero field value during inflation.
After the inflation ends, $S$ starts to roll down, and $S$ will eventually settle down at $S=\langle S\rangle$ without crossing the origin by obeying the pseudo-scaling law~\cite{Ema:2015dza}.%
\footnote{We are assuming the dynamics of $S$ is dominated by the Hubble induced mass and the sextet potential $V\sim |S|^6$.}
Therefore, the domain wall is not formed. 
We leave a further study for future works.

\bibliography{papers}

\end{document}